\documentclass[aps, twocolumn, showpacs, 10pt, superscriptaddress, floatfix]{revtex4}
\usepackage[normalem]{ulem}

\usepackage{graphicx}
\usepackage{amsmath}

\begin{document}
\title{Many-body dynamics of $p$-wave Feshbach molecule production: a mean-field approach}
\author{L. Austen}
\author{L. Cook}
\affiliation{Department of Physics and Astronomy, UCL, Gower Street, WC1E 6BT London, United Kingdom}
\author{M. D. Lee}
\affiliation{School of Mathematics, University of Southampton, Southampton SO17 1BJ, United Kingdom}
\author{J. Mur-Petit}
\email{jordi.mur@csic.es}
\altaffiliation{Present address: Instituto de Estructura de la Materia, IEM-CSIC, Serrano 123, 28006
Madrid, Spain}
\affiliation{Instituto de F\'\i sica Fundamental, IFF-CSIC, Serrano 113 bis, 28006
Madrid, Spain}


\begin{abstract}
We study the mean-field dynamics of $p$-wave Feshbach molecule production in 
an ultra-cold gas of Fermi atoms in the same internal state. We derive a
separable potential to describe the low-energy scattering properties of such
atoms, and use this potential to solve the mean-field dynamics during a magnetic
field sweep. Initially, on 
the negative scattering length side of a Feshbach resonance the gas is 
described by the BCS theory. We adapt the method by Szyma\'{n}ska et al. 
[Phys. Rev. Lett. \textbf{94}, 170402 (2005)] to $p$-wave interacting Fermi
gases and model the conversion dynamics of the gas into a Bose-Einstein 
condensate 
of molecules on the other side of the resonance under the influence of
a linearly varying magnetic field. 
We have analyzed 
the dependence of the molecule production efficiency on the density of the gas, 
temperature, initial value of the magnetic field, and magnetic field ramp speed. Our results
show that in this approximation molecule production by a linear magnetic field 
sweep is highly dependent on the initial state.
\end{abstract}
\maketitle

\section{Introduction} 

Ultra-cold Fermi gases have attracted considerable interest due to the possibility
of studying many-body fermionic physics in well controlled and tunable
environments~\cite{fermgasrev}. Some of this interest has been stimulated by the
prospect of observing pairing phenomena similar to those observed in
superconductors and liquid $^{3}$He. The existence of magnetically tunable
Feshbach resonances in ultra-cold gases~\cite{Fano_1935,Feshbach_1958,
Feshbach_1962,Tiesinga_1993,chin_RMP_2010} allows the interaction strength between
the atoms to be tuned such that the system behaves as a gas of long range Cooper
pairs described by the theory of superconductivity presented by Bardeen, Cooper
and Schrieffer (BCS)~\cite{BCSspaper} in the limit of weak interatomic
interaction, and as a Bose-Einstein condensate (BEC) of diatomic molecules in the
limit of weak repulsion~\cite{regal_nature_2003}. The transition between these
limits is often referred to as the BCS-BEC crossover~\cite{fermgasrev}.

Identical fermions in the same internal state are forbidden from colliding in the
$s$-wave ($\ell$=0, $\ell$ being the value of the relative angular momentum
quantum number) so that the lowest partial wave scattering amplitude becomes the
$p$-wave ($\ell$=1). Pairing between atoms interacting in the $p$-wave has been
theoretically studied in the context of solid-state superconductivity~\cite{Balian1963}, $^{3}$He~\cite{wolfe_1979}, particle physics~\cite{Godfrey1991} and, more recently, ultra-cold
gases~\cite{Baranov1996,Efremov2002,MurPetit2004pra,MurPetit2004jpb,Tin_lun_2005, Ohashi2005,Cheng_2005,gurarie_2005, gurarie_radzihovsky_review} . $p$-wave
Feshbach resonances have been observed in ultra-cold gases of the fermionic
isotopes $^{40}$K~\cite{regal} and $^{6}$Li~\cite{zhang, schunck}. In the case of
$^{40}$K experiments have shown the existence of a splitting of the $p$-wave
resonance which was explained by the magnetic dipole-dipole interaction between
the atomic valence electrons~\cite{ticknor}. 

The formation of weakly bound diatomic $p$-wave molecules has been 
experimentally observed in both $^{40}$K~\cite{gaebler} and 
$^{6}$Li~\cite{zhang, fuchs}. In the work on $^{40}$K, atoms were prepared in 
the state labelled $|f,m_{f} \rangle = |9/2, -7/2 \rangle$, where $f$ denotes
the hyperfine state of an atom in the absence of a magnetic field and $m_{f}$ 
denotes the Zeeman state of the atom. Molecules were formed using a resonantly
modulated magnetic field~\cite{gaebler}, as first demonstrated for 
the formation of molecules in Bose gases~\cite{Thompson_2005}. 
Binding energies of the $^{40}$K$_{2}$ molecules were measured on both
the BEC side where the atoms form a bound state and on the BCS side where the 
atoms form a resonance state localised by the centrifugal barrier. In the
latter case lifetimes for decay from the resonance state were also measured. 
Similar experiments in $^{6}$Li used linear sweeps of the magnetic field to
form molecules~\cite{fuchs,Inada_2008, Maier_2010}.

Motivated by these experiments we study the dynamics of molecule formation from a
gas of fermionic atoms in the same internal state, which occurs during a linear
sweep of the magnetic field. Such a system has been studied previously by
Szyma\'{n}ska et al.~\cite{Marzena_prl_2005}, who used a mean-field approach to
investigate $s$-wave pairing for fermions in different spin states. Here we extend
this treatment to deal with the more complicated $p$-wave pairing case. The 
initial state is taken to be a BCS paired gas on the negative scattering length
side of the resonance. We study the dependence of the molecule production
efficiency on the speed of the magnetic field sweep, as well as on temperature,
density and initial magnetic field position. We also study the evolution of the
order parameter and the molecule density after a rapid variation of the magnetic
field.

This article is structured as follows: In Sec. II, we introduce the
pseudo-potential that we use to model the two body physics in the many body
problem. In Sec. III, we present our BCS model for studying the initial state of
the gas from which to study the dynamics, while Sec. IV contains the dynamical
equations. We study linear sweeps of the magnetic field from the BCS side of the 
resonance into the BEC side where we calculate the molecule production efficiency
by taking the overlap of the pair function with the two body bound state. We also
analyze the prospects to observe atom-molecule coherent oscillations as a function
of the time after a fast sweep through the resonance. Finally, Sec. V is a
discussion of our results.

\section{Two-body model}

In order to study the many-body effects present in the formation of $p$-wave
Feshbach molecules as a linear magnetic field sweep is applied across the
resonance, we extend the methods of Szyma\'{n}ska et al.~\cite{Marzena_prl_2005}
to the case of a $p$-wave resonance. We must therefore choose a suitable
representation for the two-body physics. Due to the low energies involved in
ultra-cold collisions, it is not necessary to use the exact form of the interatomic
potential, provided that all relevant low energy scattering properties are
reproduced. Previous work on fermions has been successful in describing the
two-body physics of Feshbach resonances using pseudo-potentials in both the 
$s$-wave and $p$-wave~\cite{gurarie_radzihovsky_review}. Both the mean-field
thermodynamics and dynamics have been studied using a separable potential to model
the two-body interaction close to an $s$-wave Feshbach 
resonance~\cite{Marzena_PRA_2005,Marzena_prl_2005}. Motivated by this we seek a
similar separable form of the $p$-wave potential that can be used in the many-body
calculations.

The presence of a magnetic field defines a quantisation axis. Particles 
interacting in the $p$-wave have one quantum of relative angular momentum and 
there are three possible projections of this vector onto the quantisation axis.
The situation can be encapsulated in a three-component separable potential
\begin{equation}
 V = |\chi_{1,1} \rangle \xi_{1,1} \langle \chi_{1,1} |+ |\chi_{1,0} \rangle
 \xi_{1,0} \langle \chi_{1,0} |+ |\chi_{1,-1} \rangle \xi_{1,-1} \langle 
 \chi_{1,-1} |.
\label{full_sep_pot_form}
\end{equation}
Here $| \chi_{\ell, m_{\ell}} \rangle$ is the form factor of the potential and 
$\xi_{\ell,m_{\ell}}$ is the amplitude, where $m_{\ell}$ denotes the projection
of this angular momentum onto the magnetic field, which we choose to be
in the $z$-direction. This pseudo-potential can account for the observed
splitting of the resonance feature~\cite{ticknor} into distinct resonances
depending upon the value of $|m_\ell|$, due to magnetic dipole-dipole
interactions between the valence electrons. For each individual term a
suitable choice must be made for the form factor to recover the low energy
scattering behaviour of the atom pair. A model that reproduces the threshold behaviour
should be sufficient for our calculations since this is the energy range in
which the experiments have been performed.

We set the parameters of the pseudo-potential to reproduce physical 
observables. One feature of the two body physics that we can compare with 
experiment is the measured binding energy of the molecule. Bound states are 
associated with a pole in the $T$-matrix~\cite{Taylor} given by the 
Lippmann-Schwinger equation
\begin{equation}
T(z) = V + V G_{0}(z)T(z).
\end{equation}
Here $G_{0}(z)=(z-H_{0})^{-1}$ is the free Green's function and $H_{0}$ is the 
free Hamiltonian. Experimental measurements show
that the resonances due to the different $|m_\ell|$ values in $^{40}\mbox{K}$
are narrow and well separated~\cite{ticknor}, and so we assume each term of 
Eq.~(\ref{full_sep_pot_form}) can be treated individually leading to a
$T$-matrix for each component in the form
\begin{equation}
 T_{1 m_{1}}(z)= \frac{| \chi_{1 m_{1}} \rangle \xi_{1 m_{1}} \langle \chi_{1 m_{1}} |}{1-\xi_{1 m_{1}} \langle \chi_{1 m_{1}} | G_{0}(z) | \chi_{1 m_{1}} 
\rangle}.
\label{T_matrix}
\end{equation}
 A pole in $T_{1 m_{1}}(z)$ coincides with a vanishing denominator in 
Eq.~(\ref{T_matrix}). The $T$-matrix is related to the 
scattering amplitude through $f_{\ell,m_{\ell}}(p)=-\pi m \hbar 
\langle p \ell,m_{\ell} | T_{\ell,m_{\ell}} \left(\frac{p^{2}}{2 \mu}+i0 
\right) |p \ell,m_{\ell} \rangle$, where $p$ is the magnitude of the relative 
momentum and $m$ is the single particle mass. Given that the long range 
behaviour of the 
interatomic interaction is dominated by the van der Waals potential 
\cite{Bransden_and_joachain}, the parameter $\xi_{1 m_{1}}$ can be related to 
the scattering properties of the system from the low-energy limit of the 
partial wave scattering amplitude for the $s$-wave and $p$-wave~\cite{Newton}
\begin{equation}
\lim_{p\rightarrow 0}f_{\ell,m_{\ell}}(p) = -a_{\ell,m_{\ell}} \frac{p^{2
\ell}}{\hbar^{2 \ell}}.
\label{part_wave_scat_amp}
\end{equation}
Here $a_{\ell,m_{\ell}}$ is the partial 
wave scattering ``length'', which in the case of the $p$-wave has the 
dimensions of volume and shall here be referred to as the scattering volume. 
Furthermore, in the vicinity of a resonance the scattering volume can be 
parameterised by the magnetic field using the well known resonance formula 
\cite{Moerdijk,gubbelstoof}
\begin{equation}
 a_{\ell,m_{\ell}}(B)= a_{\ell,m_{\ell}}^{\mathrm{bg}}\left( 1 - \frac{ \Delta B_{\ell,m_{\ell}}}{B - B_{\ell,m_{\ell}}^{0}} \right).
\label{resonform}
\end{equation}
Here $a_{\ell,m_{\ell}}^{\mathrm{bg}}$ is the partial-wave background scattering 
``length'' for the $m_{\ell}$th component, $\Delta B_{\ell,m_{\ell}}$ is 
referred to as the width of the resonance and $B_{\ell,m_{\ell}}^{0}$ is the 
resonance position.
With this approach, we extend to $p$-wave interactions earlier work~\cite{Marzena_prl_2005,Marzena_PRA_2005} demonstrating that a single-channel approach with this magnetic-field dependence for the scattering length, Eq.~\eqref{resonform}, is appropriate to describe the dynamics of cold Fermi gases interacting via $s$-waves.

\begin{figure}
 \includegraphics[width=\columnwidth,clip]{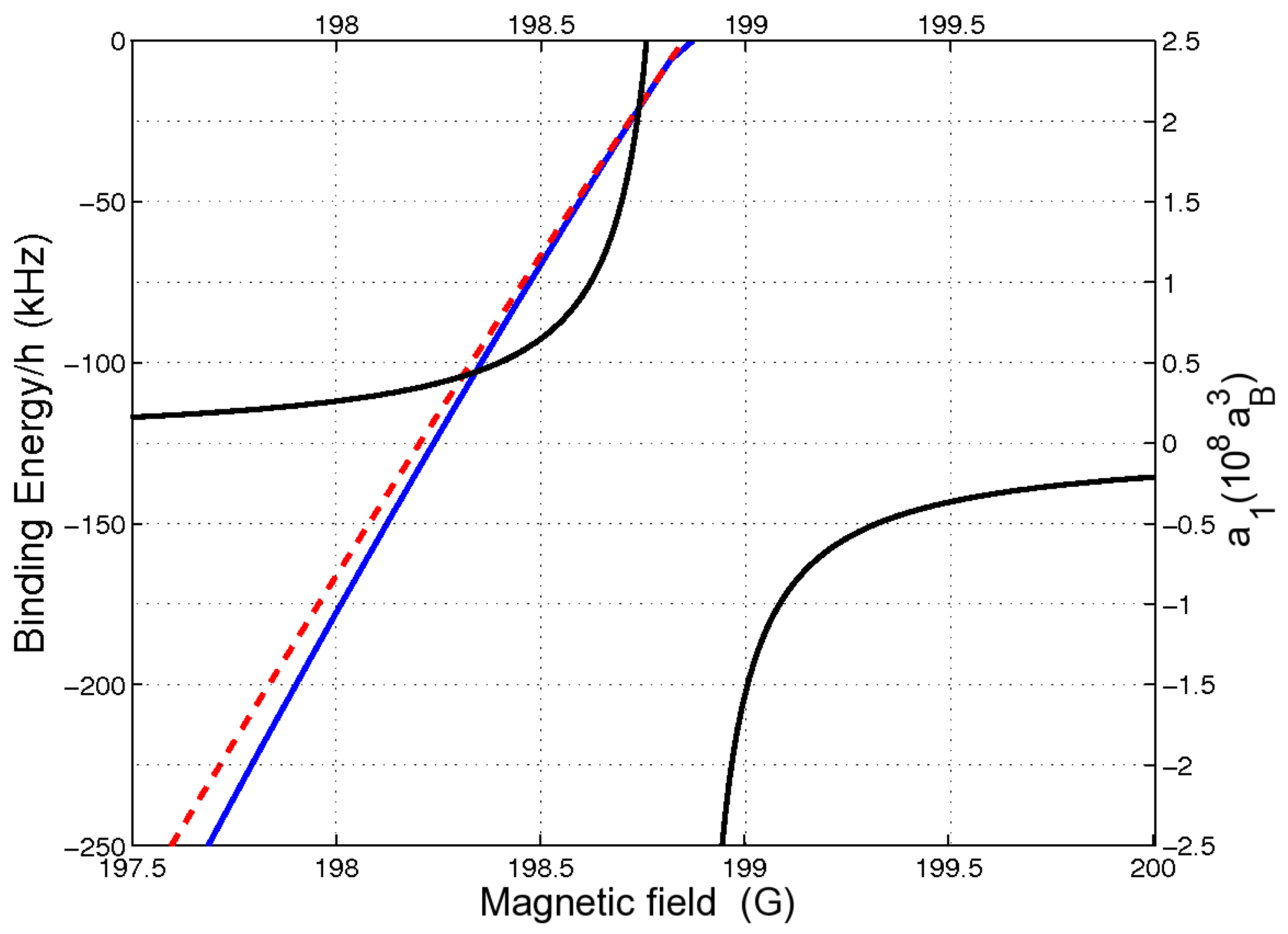}
 \caption{(Color online) Variation of the $p$-wave binding energy with the magnetic field for
the $m_{1}=0$ resonance in $^{40}K$. The solid blue line is the solution to
Eq.~(\ref{psepen}). The dashed red line is the low energy expansion of 
Eq.~(\ref{penapp}). The black solid line shows the variation of the scattering
volume (right axis) in the vicinity of the resonance.
}
 \label{penandbinfig}
\end{figure}

In Appendix~\ref{app_1} we discuss how we obtain a suitable $p$-wave form factor,
and give values for the parameters of our separable potential for the case of
$^{40}\mbox{K}$ atoms. Using this potential, we can solve
the equation for the bound state
\begin{equation}
1-\xi_{1 m_{1}} \langle \chi_{1 m_{1}} | G_{0}(E_{-1}) | \chi_{1 m_{1}} \rangle =0,
\label{gen_boun_st}
\end{equation}
where $E_{-1}$ is the energy of the least bound state of the atom pair.
We then get the low energy expansion of the bound state energy~\cite{Tin_lun_2005}
\begin{equation}
 E_{-1} \approx -\frac{\sqrt{\pi} \sigma_{1 m_{1}} \hbar^{2}}{m a_{1}},
 \label{penapp}
\end{equation}
where $\sigma_{1 m_{1}}$ is a parameter associated with the range of the potential,
as explained in Appendix~\ref{app_1}. This expression shows that the bound state
energy varies as the inverse of the scattering volume, and to this extent is in
agreement with previous two-channel calculations~\cite{Gao, Chevy, gubbelstoof}. A
plot of the bound state energy about the $B_{1,0}^{0} = $198.85~G $m_{1}=0$
resonance in $^{40}$K found from Eq.~(\ref{psepen}) is given in
Fig.~\ref{penandbinfig}. Also plotted is the low energy expansion of 
Eq.~(\ref{penapp}), which shows a good agreement close to the resonance.

\begin{figure}[t]
 \includegraphics[width=\columnwidth,clip]{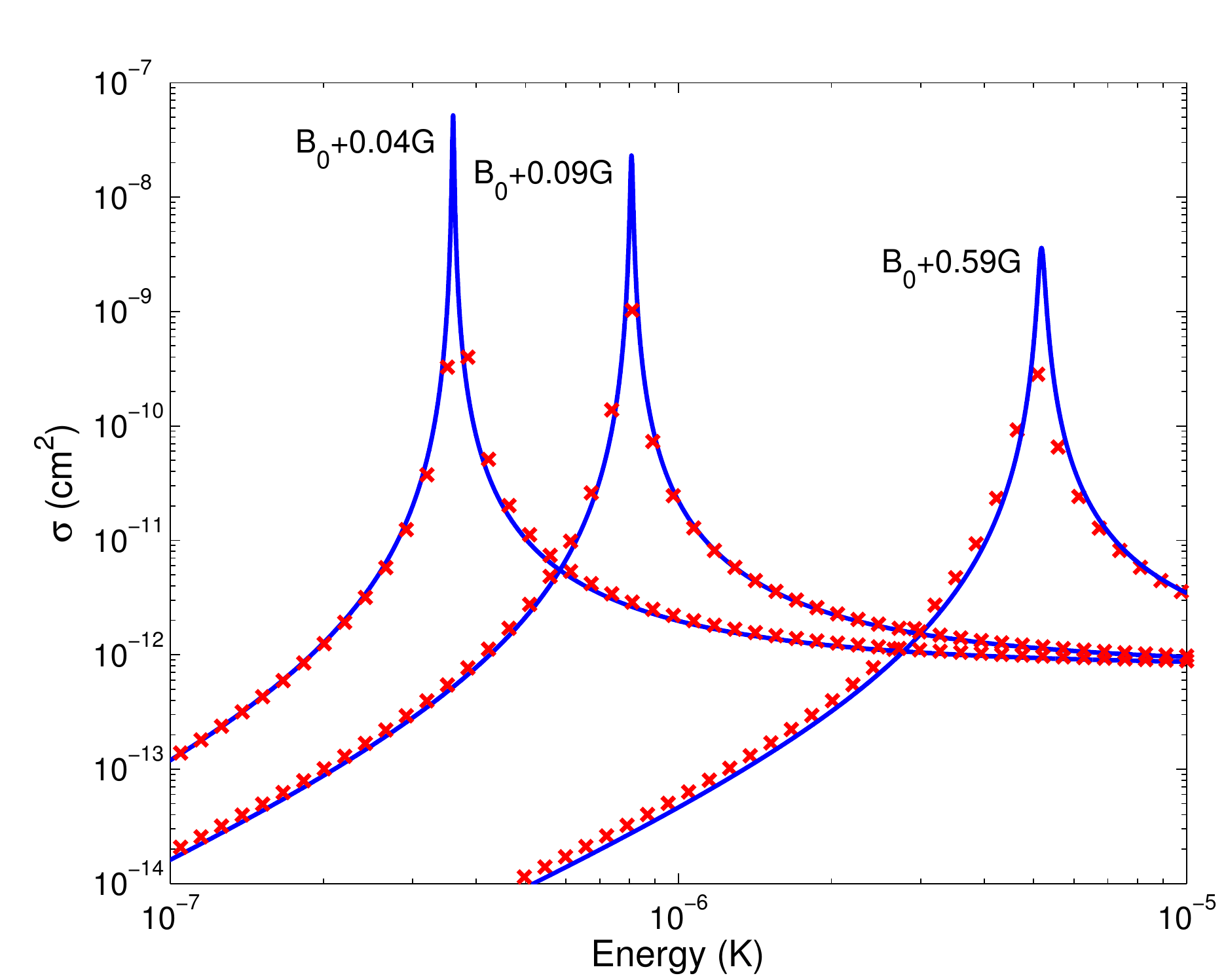}
 \caption{(Color online) $p$-wave elastic scattering cross section for $^{40}$K colliding
in the $| 9/2, -7/2 \rangle $ channel as a function of collision energy at
three different magnetic field strengths. The
solid blue line corresponds to the pseudo potential model. The red crosses
are from a coupled-channel calculation.
}
 \label{cross_compare}
\end{figure}

For $p$-wave resonances, it has been observed that the magnetic dipole-dipole
interaction plays an important role, leading for example to the splitting of the
$m_1=0$ and $m_1=\pm1$ resonances~\cite{ticknor}. For such an interaction, which
decays as $1/r^3$ at long distances, no scattering length can be properly
defined~\cite{Taylor}. However, a scattering volume model may give the correct
scattering cross section behaviour within some finite range of collision
energies. Physically this can be regarded as truncating the interaction at the
radius where it becomes small compared to the kinetic energy. To illustrate this
Fig.~\ref{cross_compare} shows the scattering cross section given by our
pseudo-potential, and compares it to a much more detailed coupled-channels
calculation. Here we obtain the coupled-channels results using the
Manolopoulos log-derivative propagation technique~\cite{manolopoulos1986}
including a dipole-dipole interaction term in the Hamiltonian~\cite{Stoof1988};
we use the Born-Oppenheimer potentials of ref.~\cite{falke08}. It can be
seen that our separable potential reproduces the low energy scattering behaviour
to a good degree over several orders of magnitude, particularly close to the
resonance, and we therefore proceed in the following sections to use this
potential in the many-body problem of molecule formation.

\section{Mean-field thermodynamics}
\label{BCS_section}

The stationary properties of a weakly attractive ultracold Fermi gas can be
described by the BCS theory of superconductivity~\cite{BCSspaper}. While trapped
gases usually have anisotropic and inhomogeneous density distributions, for
simplicity we consider here a spatially homogenous system. 

In the following we review some standard results of BCS theory~\cite{Fetter} to
present the notation we use in the following paragraphs.
The many body Hamiltonian for a system of fermions in the pairing approximation
is given by
\begin{eqnarray}
H &=& \sum_{ij} \langle i |K|j \rangle a^{\dagger}_{i}a_{j} \\\nonumber
&&+\frac{1}{2} \sum_{klmn} \langle k l | V | m n \rangle
\left( \langle a^{\dagger}_{k}a^{\dagger}_{l} \rangle a_{n}a_{m}
+ a^{\dagger}_{k}a^{\dagger}_{l} \langle a_{n}a_{m} \rangle \right).
\end{eqnarray}
Here the $a_{i}$ and $a^{\dagger}_{i}$ are the usual fermion annihilation and 
creation operators, which obey the usual fermion anti-commutation rules, $K$ is
the single particle kinetic energy operator, and $V$ is the interaction
potential. The brackets $\langle ... \rangle$ represent an average over the
thermodynamic state of the system in the grand canonical ensemble where particle
number is not fixed, $|i\rangle$ are single particle states and $|i j \rangle$ are
two particle states. The finite temperature Green's functions of the system can
be defined as
\begin{equation}
g_{rs}(\tau, \tau') = - \langle T_{\tau}[a_{r}(\tau)
a^{\dagger}_{s}(\tau')]\rangle,
\end{equation}
\begin{equation}
F^{\dagger}_{rs}(\tau,\tau') = - \langle T_{\tau}[a^{\dagger}_{r}(\tau)
a^{\dagger}_{s}(\tau')]\rangle,
\end{equation}
the latter representing pairing in the gas. Here $T_{\tau}$ is the imaginary
time ordering operator that puts the smallest value of $\tau$ to right 
\cite{Fetter}. It is also often convenient to define a gap function
\begin{equation}
\Delta_{rs} = \sum_{ij} \langle a_{i}a_{j} \rangle \langle r s | V |ij \rangle,
\label{gapdef}
\end{equation}
which acts as the order parameter 
of our system.

By working in the momentum representation and considering a translationally 
invariant system we can write down the Heisenberg equations of motion for the
Green's functions. The resulting equations can be Fourier transformed and combined with 
Eq.~(\ref{gapdef}) to give
\begin{equation}
\Delta^{*} (\mathbf{p}) = - \int d^{3} q \; \langle \mathbf{p} | V | \mathbf{q} \rangle
\frac{ \Delta^{*}(\mathbf{q})}{2 \epsilon_{\mathbf{q}}} \tanh(\beta 
\frac{\epsilon_{\mathbf{q}}}{2}),
\label{gapequation_text}
\end{equation}
and
\begin{equation}
\frac{N}{{\cal V}}=\frac{1}{2}\int d^{3} q \left[ 1 -
\frac{E_{q}}{2 \epsilon_{\mathbf{q}}} \tanh(\beta 
\frac{\epsilon_{\mathbf{q}}}{2}) \right],
\label{densityequation}
\end{equation}
where $\epsilon_{
\mathbf{q}}=\left(E_{q}^{2}+|\Delta(\mathbf{q})|^{2}\right)^{1/2}$,
$E_{q}=\frac{q^{2}}{2m}-\mu$, ${\cal V}$ is the volume of the system, $\beta=\frac{1}{k_{\mathrm{B}}T}$ and $k_{\mathrm{B}}$ is 
Boltzmann's constant. Equations~(\ref{gapequation_text}) 
and~(\ref{densityequation}) are often referred to as the gap 
equation and the density equation, respectively. These are the BCS equations for 
the system and must be solved simultaneously.
Experiments are usually performed at constant temperature, which fixes
$\beta$ in our system. This leaves the chemical potential, $\mu$, and 
$\Delta(p)$, as parameters to be solved for.

Using the potential of Eq.~(\ref{full_sep_pot_form}) the order parameter can 
be written as
\begin{equation}
\Delta^{*}(\mathbf{p})=\sum_{m_{1}}\Delta^{*}_{m_{1}} \chi_{m_{1}}(p) Y^{*}_{1,m_{1}}(\hat{\mathbf{p}}), 
\end{equation}
where $\Delta_{m_{1}}$ and $\chi_{m_{1}}(p) = \langle p 1,m_{\ell} | \chi_{1,m_{\ell}} \rangle$ represent components in the 
$p$-wave. We now drop the label $\ell$ in the subindex in the functions we 
have defined, since we will only be concerned with the $p$-wave, $\ell=1$.
By inserting this into Eq.~(\ref{gapequation_text}) we arrive at a set of coupled 
equations for all components of the gap in the angular momentum basis
\begin{widetext}
\begin{equation} 
\label{BCS_three_components}
\Delta^{*}_{m_{1}} = - \int d^{3}q \sum_{m_{1}'} \frac{\chi_{m_{1}}(q)
Y_{1,m_{1}}(\hat{\mathbf{q}}) \xi_{m_{1}} \Delta^{*}_{m_{1}'} 
\chi_{m_{1}'}(q)Y^{*}_{1,m_{1}'}(\hat{\mathbf{q}})}
{2 \epsilon_{\mathbf{q}}} 
\tanh \left[
\frac{ \beta}{2} \epsilon_{\mathbf{q}} \right].
\end{equation}
\end{widetext}
Experiments have shown the $m_{1}=1$ and $m_{1}=-1$ states to be 
degenerate~\cite{ticknor}. The degeneracy of the two components is due 
to the rotational symmetry of the system about the magnetic field axis. 
The parameter $\Delta_{m_{1}}$ for the $m_{1}=1$ and $m_{1}=-1$ components can 
be written as
\begin{equation}
\Delta_{\pm 1} = \mp \frac{1}{\sqrt{2}} \left(\Delta_{x} \pm i \Delta_{y} 
\right),
\end{equation}
where the magnitudes of these components are identical. 

It is possible to reduce the set of three coupled nonlinear equations to 
two~\cite{Luke_Austen_thesis}, so that we can write
\begin{eqnarray}
\left(\begin{array}{c}\Delta^{*}_{x} \\ \sqrt{\frac{\xi_{x}}{\xi_{z}}}\Delta^{*}_{z} \end{array}
\right)&=&
\left(\begin{array}{cc} \langle \psi_{x}(\mathbf{q}),\psi_{x}(\mathbf{q}) \rangle & 
\langle \psi_{x}(\mathbf{q}),\psi_{z}(\mathbf{q}) \rangle\\ \langle \psi_{z}(\mathbf{q})
,\psi_{x}(\mathbf{q}) \rangle & \langle \psi_{z}(\mathbf{q}),\psi_{z}(\mathbf{q} \rangle
\end{array} \right) \nonumber \\
&&\times\left(\begin{array}{c}\Delta^{*}_{x} \\\sqrt{\frac{\xi_{x}}{\xi_{z}}}\Delta^{*}_{z} \end{array}
\right).
\label{cates_matrix}
\end{eqnarray}
Here brackets $\langle \cdot , \cdot \rangle$ represent scalar products of two functions and should not be 
confused with a thermodynamic average. The functions are given by,
\begin{equation}
\psi_{x}(\mathbf{q}) = i \sqrt{\frac{3 f(\mathbf{q}) \xi_{x}}{4 \pi}} \chi_{11}(q)
 \sin \theta
\cos \phi,
\end{equation}
and
\begin{equation}
\psi_{z}(\mathbf{q}) = i \sqrt{\frac{3 f(\mathbf{q}) \xi_{z}}{4 \pi}} \chi_{0}(q)
\cos \theta,
\end{equation}
where for brevity we have defined 
\begin{equation}
f \left(\mathbf{q}\right)=\frac{\tanh \left[
\frac{ \beta}{2} \epsilon_{\mathbf{q}} \right]}{ \epsilon_{\mathbf{q}}}.
\end{equation}

\begin{figure}[t]
 \includegraphics[width=\columnwidth,clip]{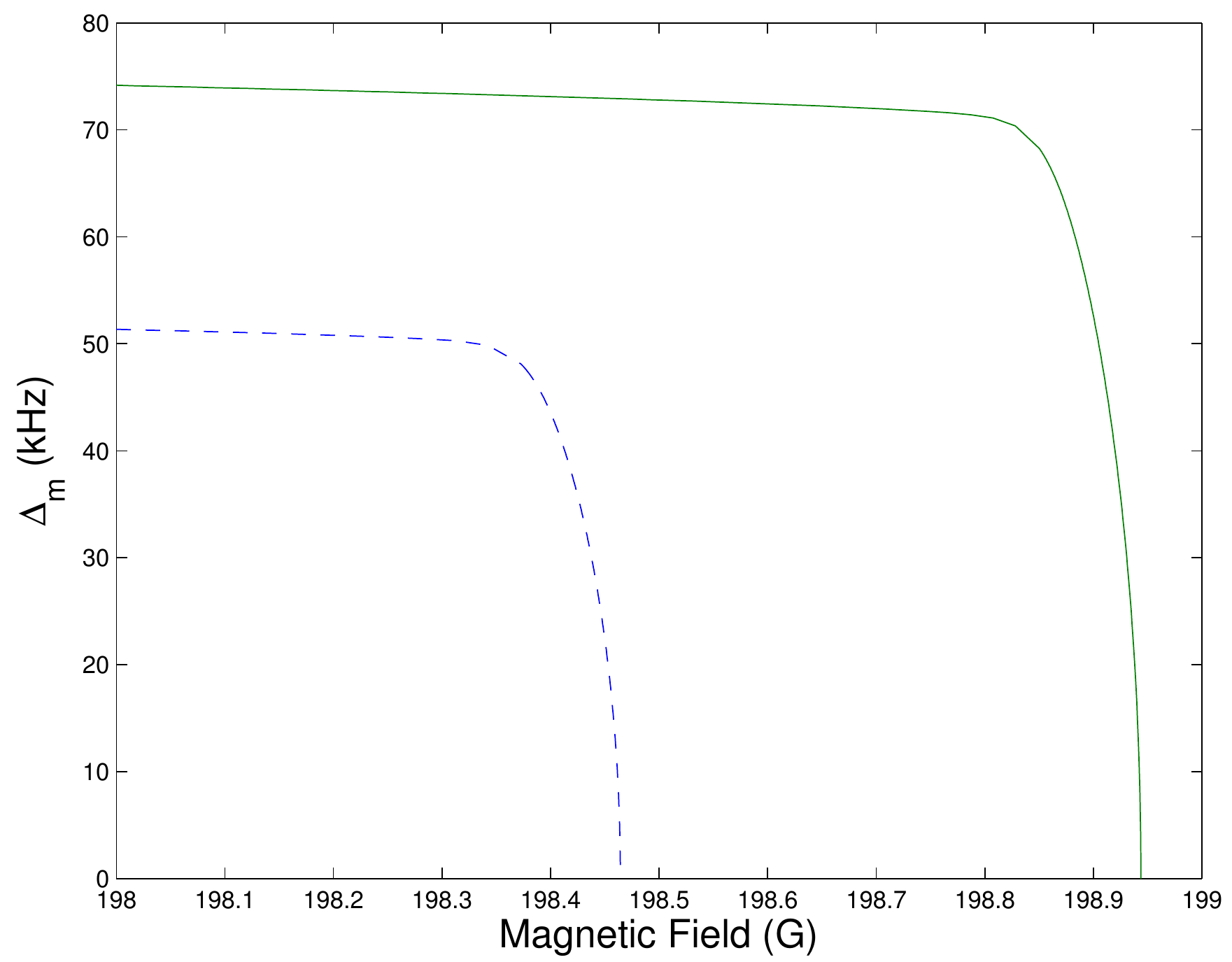}
 \caption{(Color online) Variation of the parameter $\Delta_{m}$ with magnetic field for
 the $p$-wave resonance in $^{40}\mbox{K}$ for a density of
 $10^{13}\mbox{cm}^{-3}$ and a 
temperature of $70\mbox{nK}$. The solid green line is the value of the $m=0$ 
resonance and the dashed blue line is for the $|m|=1$ resonance. There is no
 significant difference between the results obtained when coupling between 
the components is included and when the coupling is excluded.
}
 \label{gap_m_1_and_m_0}
\end{figure}

\begin{figure}[t]
 \includegraphics[width=\columnwidth,clip]{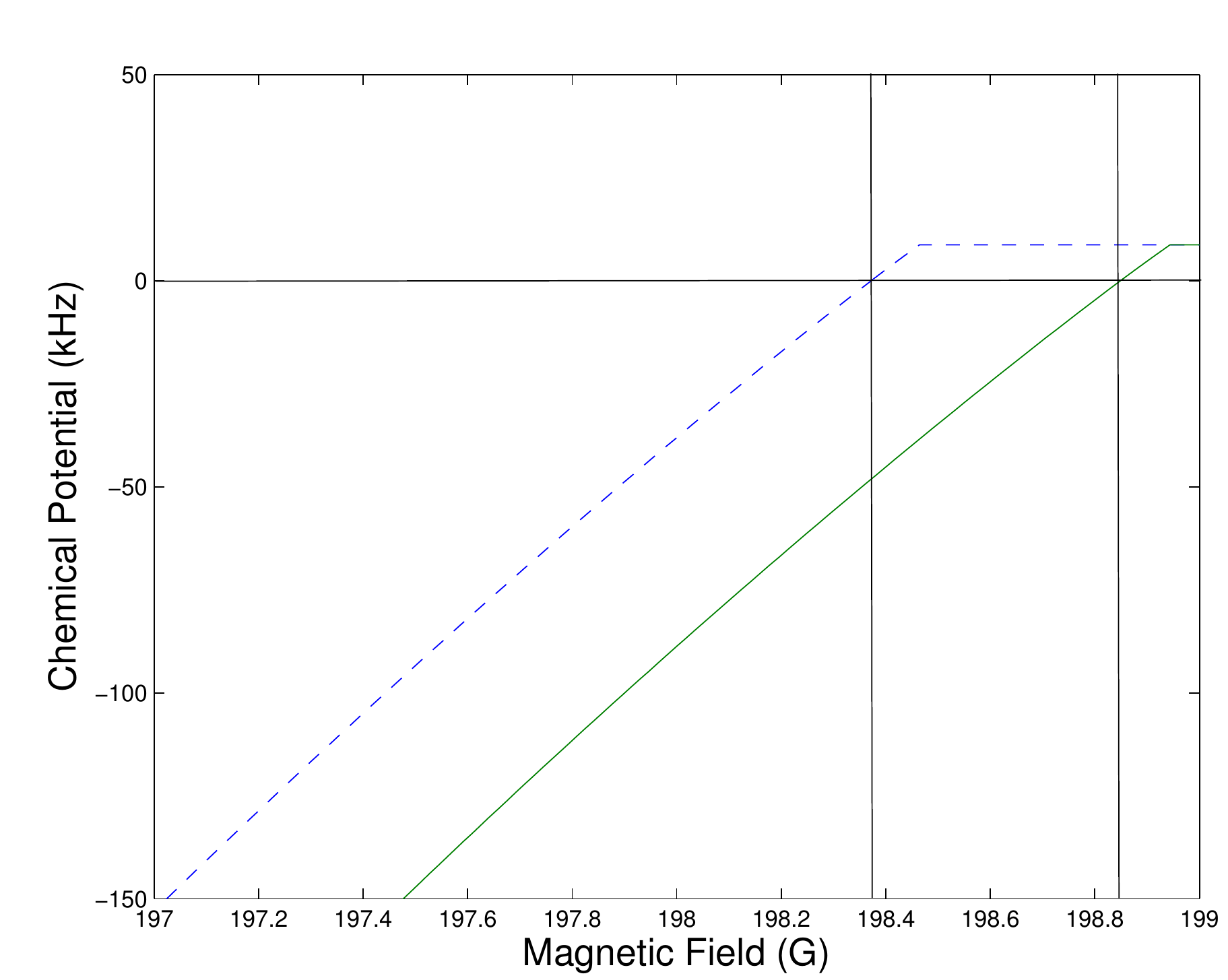}
 \caption{(Color online) Values of the $p$-wave chemical potential for the resonance 
in $^{40}\mbox{K}$ for a density of $10^{13}$cm$^{-3}$ and a temperature of $70\mbox{nK}$. The 
dashed blue line is the $|m_{1}|=1$ resonance and the solid green line is 
the $m_{1}=0$ resonance. The values for coupled resonances and separated 
resonances are indistinguishable. The positions of the resonances are marked 
by the vertical lines.
}
\label{chempot_m_1_and_m_0}
\end{figure}

We have solved Eq.~(\ref{cates_matrix}) for the resonance around 198.85~G in 
$^{40}$K. We 
found that the off-diagonal matrix elements are several orders of magnitude
smaller than the diagonal elements for densities of $10^{13}$~cm$^{-3}$ to 
$10^{15}$~cm$^{-3}$. In this case neglecting the off-diagonal elements does not
noticeably change the solution. A plot of the solutions to the BCS equations 
for $^{40}$K is shown in Fig.~\ref{gap_m_1_and_m_0}, for the gap parameters, $\Delta_{m_{1}}$ and 
Fig.~\ref{chempot_m_1_and_m_0} for the chemical potentials $\mu$.
Previous work has used the Bose-Fermi~\cite{Holland_2001, Timmermans2001228}
model to study the thermodynamic properties of resonant $p$-wave gases 
\cite{Tin_lun_2005, Cheng_2005, levinson_2007, gurarie_2005, gurarie_radzihovsky_review}. We have calculated the chemical potential and gap parameters using
both models and shown the results of the two models to be very similar 
throughout the crossover region~\cite{Luke_Austen_thesis}.

\section{Mean-field dynamics}
\subsection{Dynamics of molecule production}

Experiments on the formation of molecules often use finite speed magnetic field variations to create molecules
from gases of ultra-cold fermions 
\cite{regal_nature_2003, Strecker_2003, Cubizoles_2003, Jochim_2003}. This
has stimulated theoretical research into studying the dynamics of molecule
production in these gases 
\cite{Barankov_2004, Barankov_2004_2,Andreev_2004,altman_2005, Marzena_prl_2005,Dobrescu_2006,Tikhonenkov_2006, matyjaskiewicz_2008, gurarie_2009}.
At the mean-field level the gas can be described by two correlation 
functions: the pair function, $\Phi_{ij}(t) = \langle a_{j} a_{i} \rangle_{t}$, 
and the one body 
density matrix $\Gamma_{ij}(t)=\langle a^{\dagger}_{j} a_{i} \rangle_{t}$. 
The brackets $\langle ... \rangle_{t}$ represent a thermal average at time $t$.
Their Heisenberg equations of motion in the momentum representation for a 
homogeneous system are~\cite{Marzena_prl_2005}
\begin{align}
 i \hbar \frac{\partial}{\partial t} \Gamma (\mathbf{p}, t)
 &= 2(2 \pi \hbar)^{3/2} i \; \mathrm{Im} \left( \Phi^{*} (\mathbf{p},t) \langle \mathbf{p} | V | \Phi (t) \rangle
 \right)
 \label{dynamic_dens_eq}
 \\
 i \hbar \frac{\partial}{\partial t} \Phi (\mathbf{p},t)
 &= \langle \mathbf{p} | H_{\mathrm{2B}} | \Phi (t) \rangle
 \nonumber
 \\
 &- \langle \mathbf{p} | V | \Phi (t) \rangle \Gamma (-\mathbf{p},t)(2 \pi \hbar)^{3/2}
 \nonumber
 \\
 &- \langle \mathbf{p} | V | \Phi (t) \rangle \Gamma (\mathbf{p},t)(2 \pi \hbar)^{3/2}.
 \label{dynamic_pair_eq}
\end{align}
where due to translational invariance we have
$\Gamma_{\mathbf{p},\mathbf{p}} = \Gamma(\mathbf{p})\delta(0)$ and
$\Phi_{\mathbf{p},-\mathbf{p}} = \Phi(\mathbf{p})\delta(0)$.
The initial correlation functions are related to the solutions of the 
BCS equations of Sect.~\ref{BCS_section} through
\begin{equation}
\frac{N}{{\cal V}} = \int \frac{d^{3}p}{\left(2 \pi \hbar \right)^{3/2}} \; \Gamma 
(\mathbf{p})
\end{equation}
and
\begin{equation}
\Delta(q) = \int d^{3} q \; \langle \mathbf{p}|V| \mathbf{q} \rangle
\Phi(\mathbf{q}) 
\end{equation}
These equations form a closed set of equations for the
density distributions, without need for auxilary parameters or
constraints~\cite{Marzena_prl_2005}. Dynamical equations can be written for the 
partial wave components of both the pair function and the one-body density 
matrix. For example, the component of the pair function is given by 
\begin{equation}
\Phi_{\ell,m_{\ell}}(q,t) = i^{\ell} \int d \Omega \; Y^{*}_{\ell,m_{\ell}}
\left( \Omega \right) \Phi (\mathbf{q}, t),
\end{equation} 
where $\Omega$ is the solid angle in $\mathbf{q}$. A similar expression can be
written for the density matrix components, $\Gamma_{\ell,m_{\ell}}(q,t)$.
This allows the angular 
integrals in the dynamic equations to be performed analytically, reducing the 
dimensionality of the numerical problem provided the contribution from higher 
partial waves is not significant. When the initial conditions allow the 
resonances to be treated separately the initial state of the gas is 
in either an $|m_{1}|=1$ state or an $m_{1}=0$ state. At this level of 
approximation it is not possible to transfer population between these two 
components through the dynamics and we therefore treat them as individual cases.
We study the case of a linear magnetic field sweep
\begin{equation}
B(t) = B_{\mathrm{i}} - \dot{B}t
\label{mag_field_var}
\end{equation}
where $B_{\mathrm{i}}$ is the initial magnetic field above the resonance, $t=0$ 
is the time at which the initial state is prepared and $\dot{B}$ is the sweep 
rate. At some point the sweep 
will cease to produce molecules, so that we may choose a
final magnetic field in the BEC side of the resonance where the molecule 
production has saturated. 

The density of molecules at the end of the sweep 
can be calculated from a Wick expansion of the two-body correlation function. 
This can be approximately
written as an overlap with the two-body bound state~\cite{Marzena_prl_2005}
 \begin{equation}
n_{\mathrm{mol}} = \frac{1}{2}\left| \int d^{3} p \; \phi^{*}_{\mathrm{b}}(\mathbf{p}) \Phi (\mathbf{p})\right|^{2},
\label{dens_of_mols}
\end{equation}
where $\phi^{*}_{\mathrm{b}}(\mathbf{p})$ is the two-body bound state wavefunction in the 
momentum representation obtained by solving the Schr\"{o}dinger equation using
the separable potential. In Eq.~(\ref{dens_of_mols}) the pair function, 
$\Phi(\mathbf{p})$ is evaluated at the final magnetic field strength of the sweep.

\begin{figure}
 \includegraphics[width=\columnwidth,clip]{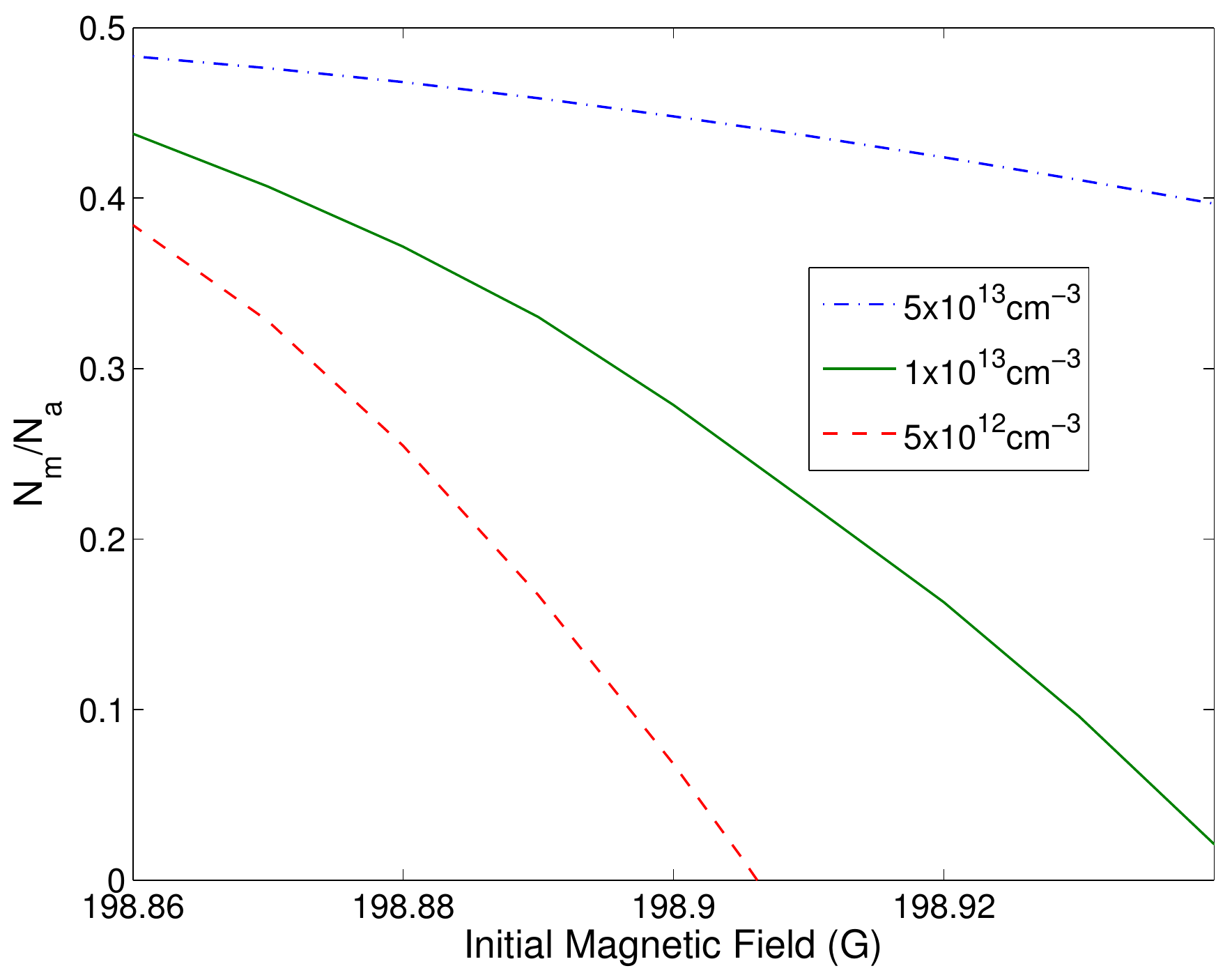}
\caption{(Color online) Variation of the molecular production efficiency in
$^{40}\mbox{K}$ as the 
initial value of the magnetic field is varied. The different lines represent 
different densities. Higher densities give a larger production efficiency. 
The temperature is $70$nK and the sweep rate is held constant at $\dot{B}=60$
G/ms. 
}
\label{B_i_vary_plot}
\end{figure}

As a first approximation we retain the lowest order partial waves. The 
effect of adding in the higher partial waves can then be investigated. To 
study the dynamics of $m_{1}=0$ molecule formation in this approximation we retain 
the functions $\Phi_{10}(p,t)$ and $\Gamma_{00}(p,t)$.

\begin{figure}[t]
 \includegraphics[width=\columnwidth,clip]{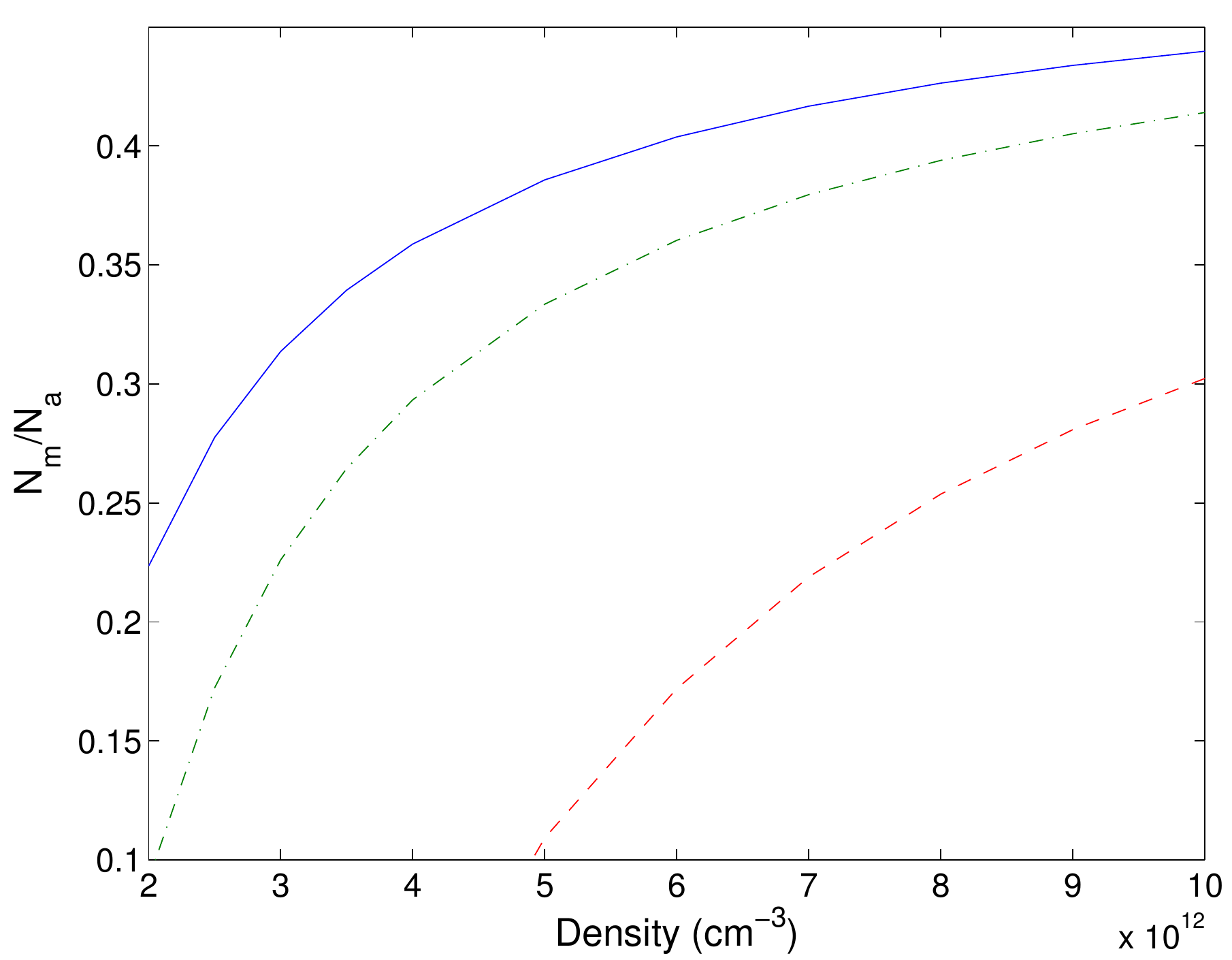}
\caption{(Color online) Variation of the molecular production in $^{40}\mbox{K}$ efficiency as the 
atomic density is varied. The sweep rate is held constant at $\dot{B}=60$ G/ms. The lines
represent temperatures of 70nK (solid blue), 100nK (dot-dashed green) and
200nK (dashed red).
}
\label{den_vary_plot}
\end{figure}

We have investigated the variation in $m_{1}=0$ molecule production as a 
function of the initial magnetic field, $B_{\mathrm{i}}$. In this case the 
temperature, atomic density and ramp speed of the magnetic field are held 
constant. The result is plotted in Fig.~\ref{B_i_vary_plot}. As the initial
magnetic field moves closer to the the resonance position the molecule 
production increases. This is due to an increase in the pairing in the initial 
state of the gas, as can be seen from Fig.~\ref{gap_m_1_and_m_0}.
Similarly, increasing the atomic density increases the 
production efficiency as shown in Fig.~\ref{den_vary_plot}. We remark that our
mean-field calculations predict a saturation of the molecule production
efficiency at increasing density. This is an expected many-body effect, in contrast to purely
two-body calculations that would predict $N_m/N_a$ to be linearly proportional
to the density.

Investigating the variation of the production efficiency as a function of initial
magnetic field proved to be difficult at low densities~($\sim 10^{12} -
10^{13}$~cm$^{-3}$). This is because the gap parameter quickly drops to zero on
the BCS side of the resonance at these low densities. At such densities it should
then be possible to approximate the production efficiency at the mean-field level
with an instant projection of the initial pair state onto the bound state. This is
because when the initial field is close to the resonance the quantities of interest
will not evolve significantly before the final magnetic field is reached.

\begin{figure}[t]
\includegraphics[width=\columnwidth,clip]{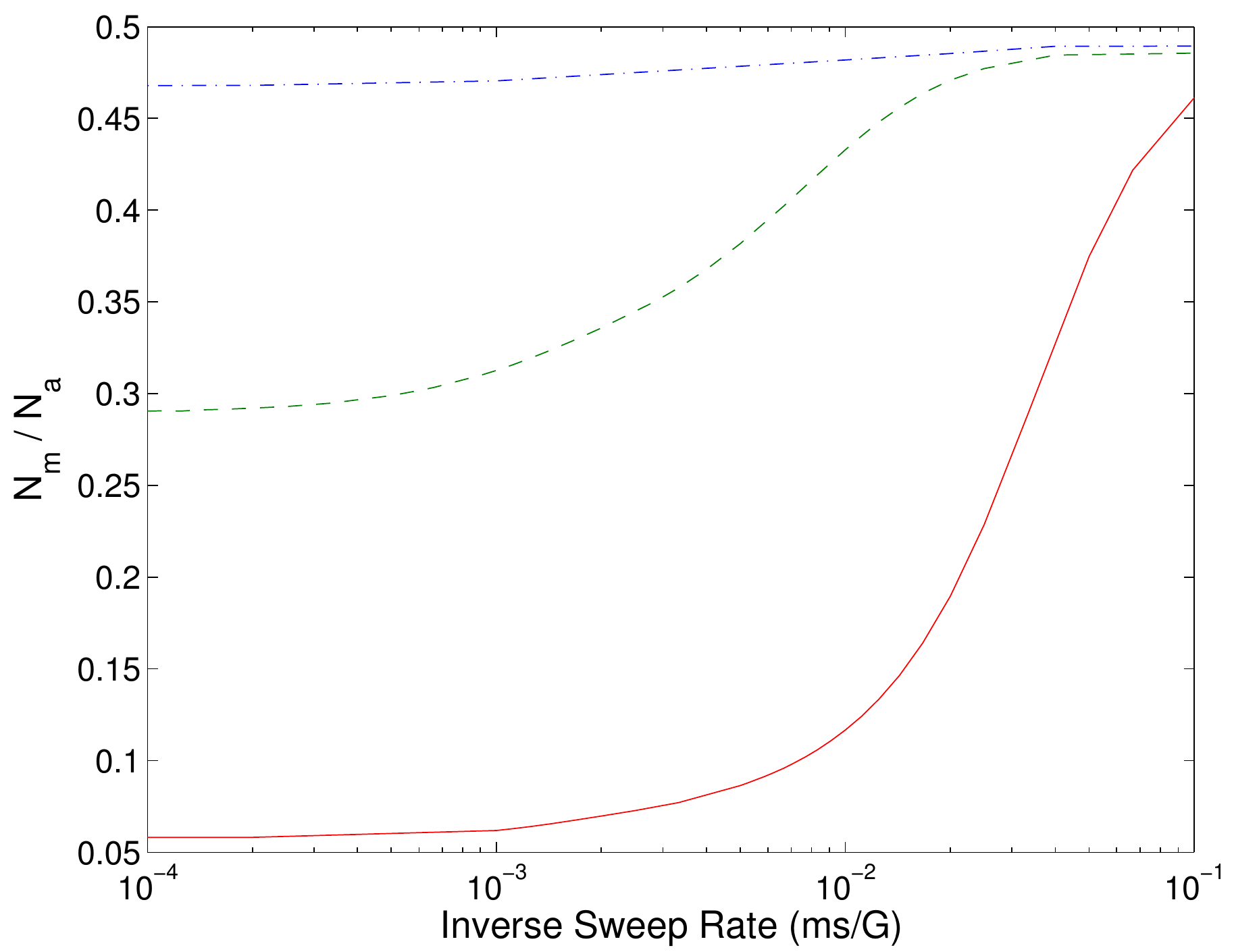}
 \caption{(Color online) Variation of the molecular production efficiency in $^{40}K$ as the 
 ramp speed is varied for a temperature of 70nK. The blue, dot-dashed curve
refers to an atomic density of $10^{15}$~cm$^{-3}$ and an initial magnetic field of 199~G.
 The green, dashed curve
refers to an atomic density of $10^{15}$~cm$^{-3}$ and an initial magnetic field of 200~G.
The red, solid curve
refers to an atomic density of $4 \times 10^{14}$~cm$^{-3}$ and an initial magnetic field of 200~G.}
\label{fig_7_8_combo}
\end{figure}

It has been shown by Iskin et al.~\cite{Iskin_2008} that for a harmonically
trapped $p$-wave superfluid the central density is much larger than for the
corresponding $s$-wave superfluid. This indicates that it may be feasible to
reach higher densities in $p$-wave Fermi gases. Following this observation, we
performed calculations setting the initial atomic density to a value well above
those normally used in experiments~(greater than $ 10^{14}$~cm$^{-3}$). In
Fig.~\ref{fig_7_8_combo} we present calculations performed for a initial atomic
density of $10^{15}~\mbox{cm}^{-3}$. These results show that it is possible to
produce a significant number of molecules even when pairing in the initial state of
the gas is not very large. This can be achieved by adjusting the sweep rate to
lower and lower values, thus allowing for two-body correlations to build up during
the sweep via intrinsically many-body processes. The closer the value of
$B_{\mathrm{i}}$ is to the resonance the more difficult it is to influence the number
of molecules produced by varying the sweep rate. This is because there is already
a great amount of pairing in the gas and the production efficiency $N_m/N_a$ can not
exceed $0.5$. On the other hand, at low ramp speeds the production efficiencies
all converge and become independent of $B_{\mathrm{i}}$, following the adiabatic
value obtained from the thermodynamic state.

It may still be possible to produce molecules from a linear sweep of the magnetic
field at lower densities. As a comparison, the molecule production at atomic
densities of $4 \times 10^{14}\mbox{cm}^{-3}$ as a function of inverse sweep rate
has been plotted in Fig.~\ref{fig_7_8_combo} (solid, red line). The initial
magnetic field has been set to $200\mbox{G}$ so that a direct comparison can be made with
the dashed-green line. More molecules are produced relative to the production from
an infinitely fast sweep for the lower density of $4 \times
10^{14}\mbox{cm}^{-3}$. However, to achieve a similar net production efficiency
as the $10^{15}\mbox{cm}^{-3}$ gas, the ramp speed will have to be set to
significantly lower values where effects due to higher order correlation functions
(not captured in the present mean-field approach) may become relevant.

\begin{figure}
 \includegraphics[width=\columnwidth,clip]{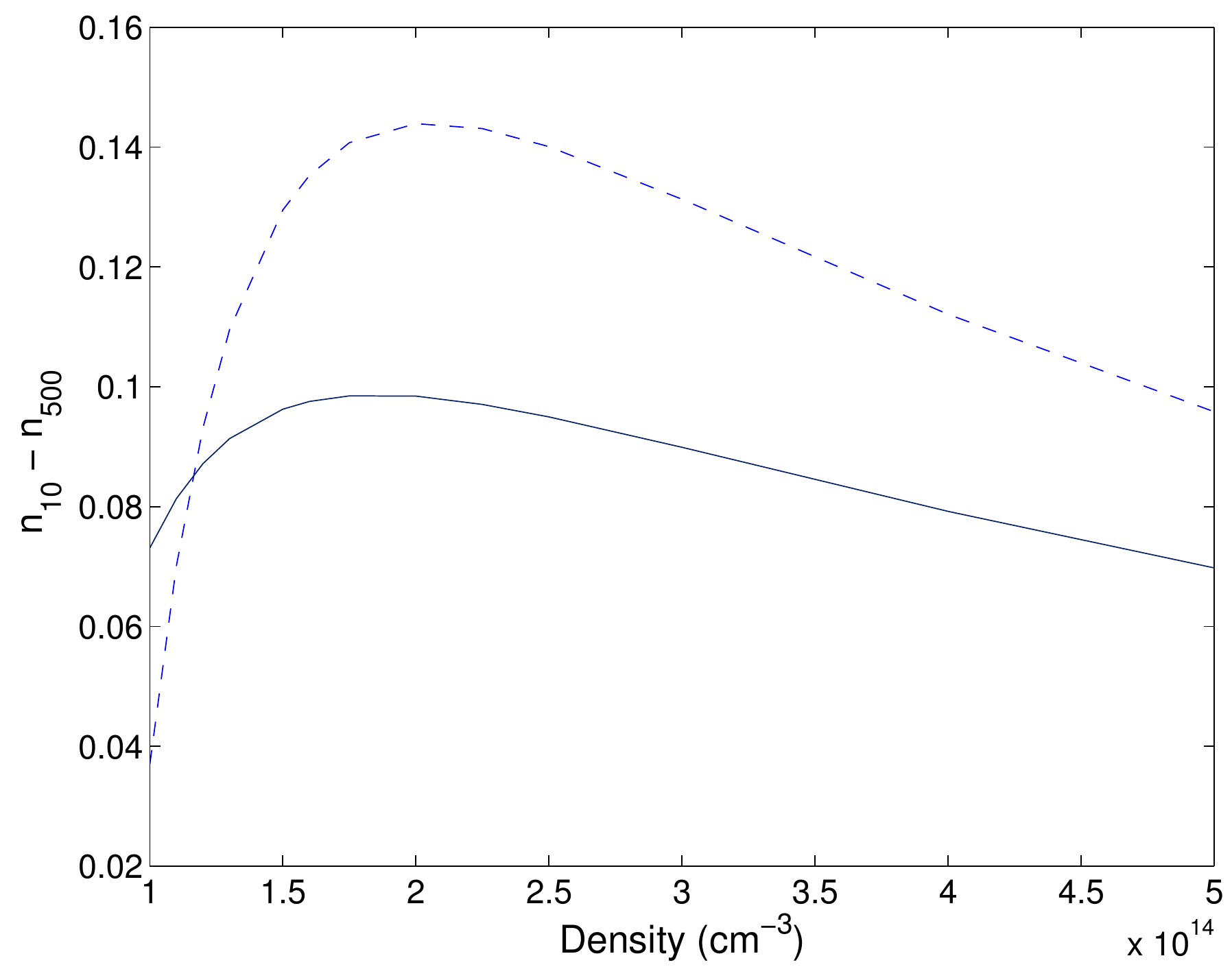}
\caption{The difference in the molecule production efficiency from a sweep
of $10 \mbox{G/ms}$ and a sweep of $500 \mbox{G/ms}$ as a function of density. Here, 
$n_{\dot{B}}$ is the number of molecules over the number of atoms after a sweep
at a speed equal to $\dot{B}$ in G/ms for a temperature of 70nK. This shows 
how many molecules are actually produced during the dynamics. The dashed curve
refers to an initial magnetic field position of $B_{i}=199.3\mbox{G}$, while the lower
curve refers to $B_{i}=199.2\mbox{G}$. It can be seen there is an optimum density
at which to produce molecules from the dynamics. 
}
\label{den_vary_10m100}
\end{figure}

To analyze how many molecules are actually produced during the dynamics, we show in
Fig.~\ref{den_vary_10m100} the fraction of molecules produced from a
$500\mbox{G/ms}$ sweep subtracted from the fraction produced from a
$10\mbox{G/ms}$ sweep. It can be seen that there is a maximum in the production
efficiency at a certain value of the density. This peak increases in magnitude
and moves to higher densities as the initial value of the magnetic field $B_{i}$
is moved to higher fields. This shows that combinations of initial density and
initial magnetic field can be picked to optimise the production efficiency.

We have checked that including higher partial wave components to study
the dynamics does not significantly affect the production
efficiencies reported here~\cite{Luke_Austen_thesis}.
It appears that retaining only the lowest $\ell$ components is sufficient to calculate the production at the mean-field level.
Similar results are obtained for the $m_{1}=1$ molecules by allowing
the initial state of the gas to be paired with $m_{1}=1$ symmetry.

\subsection{Atom-molecule coherence}

In experiments on $s$-wave molecule production, rapid sweeps of the magnetic field were
used to probe the state of the Fermi gas in the region about the resonance. It
was hypothesised that if the magnetic field was swept into the BEC side fast 
enough, such that the typical sweep time was less than the typical collision 
time, then it would be possible to extract information about the gas in the 
strongly interacting region~\cite{Regal_2004}. The question then arises of how
the state evolves after such a sweep. If the final state, held at a fixed field
value, undergoes processes that significantly change it, then this method may 
not be a reliable way of probing the gas. For the $s$-wave it has been shown 
that under such a magnetic field variation the final molecule production 
efficiency will
oscillate but with a small, decreasing amplitude~\cite{Marzena_prl_2005}. We 
use an essentially identical method to show that this is also true in the 
$p$-wave and it would not be possible to observe atom-molecule coherence
with this approach.

\begin{figure}[t]
\centering
 \includegraphics[width=\columnwidth,clip]{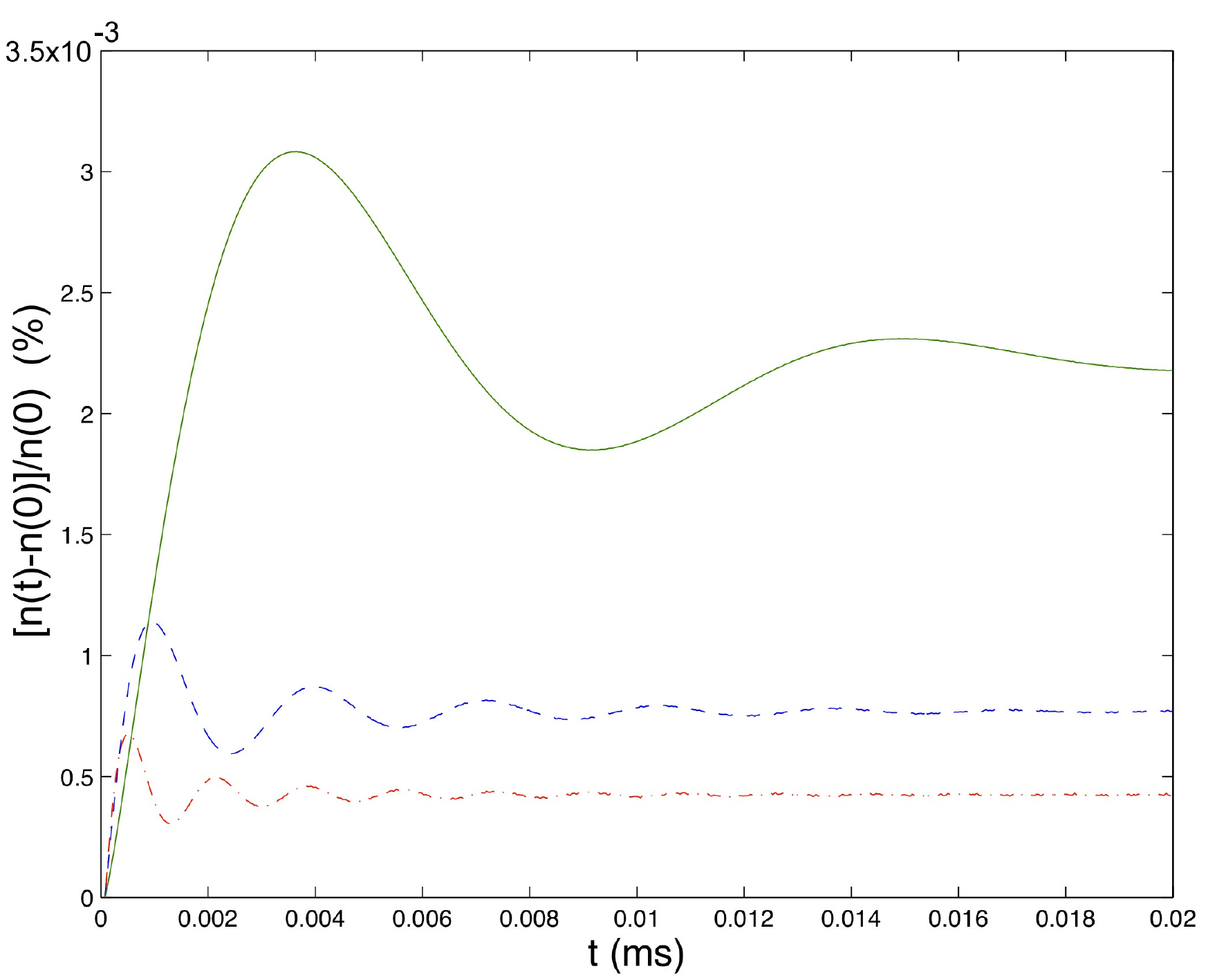}
\caption{(Color online) Evolution of the molecule production efficiency after an infinitely 
fast sweep of the magnetic field across the 198.85~G resonance in $^{40}$K. The
initial magnetic field is 199~G, just above the resonance. The different lines
correspond to differing final magnetic fields of 198.5~G (solid, green line),
197.5~G (dashed, blue line) and 196.5~G (dot-dashed, red line). $n(t)$ is the
density of molecules as a function of time where $n(0)$ is the density of 
molecules directly after the magnetic field variation.}
\label{dens_osc_varybf}
\end{figure}

Fig.~\ref{dens_osc_varybf} shows the variation in the production efficiency
as a function of time after such a magnetic field variation. In this figure, 
the
different lines correspond to different final magnetic fields. The variation
in the molecule production over this time period is given as a percentage and
seen to be on the order of $10^{-3}~\%$, which is very small. The oscillations
in the production are heavily damped with the oscillation frequency and damping rate increasing as the final magnetic field moves away from the 
resonance. For the case where the final field is located at 196.5~G the
oscillations are not visible on the scale of the figure after 20 $\mu$s.

\begin{figure}[th]
\centering
 \includegraphics[width=\columnwidth,clip]{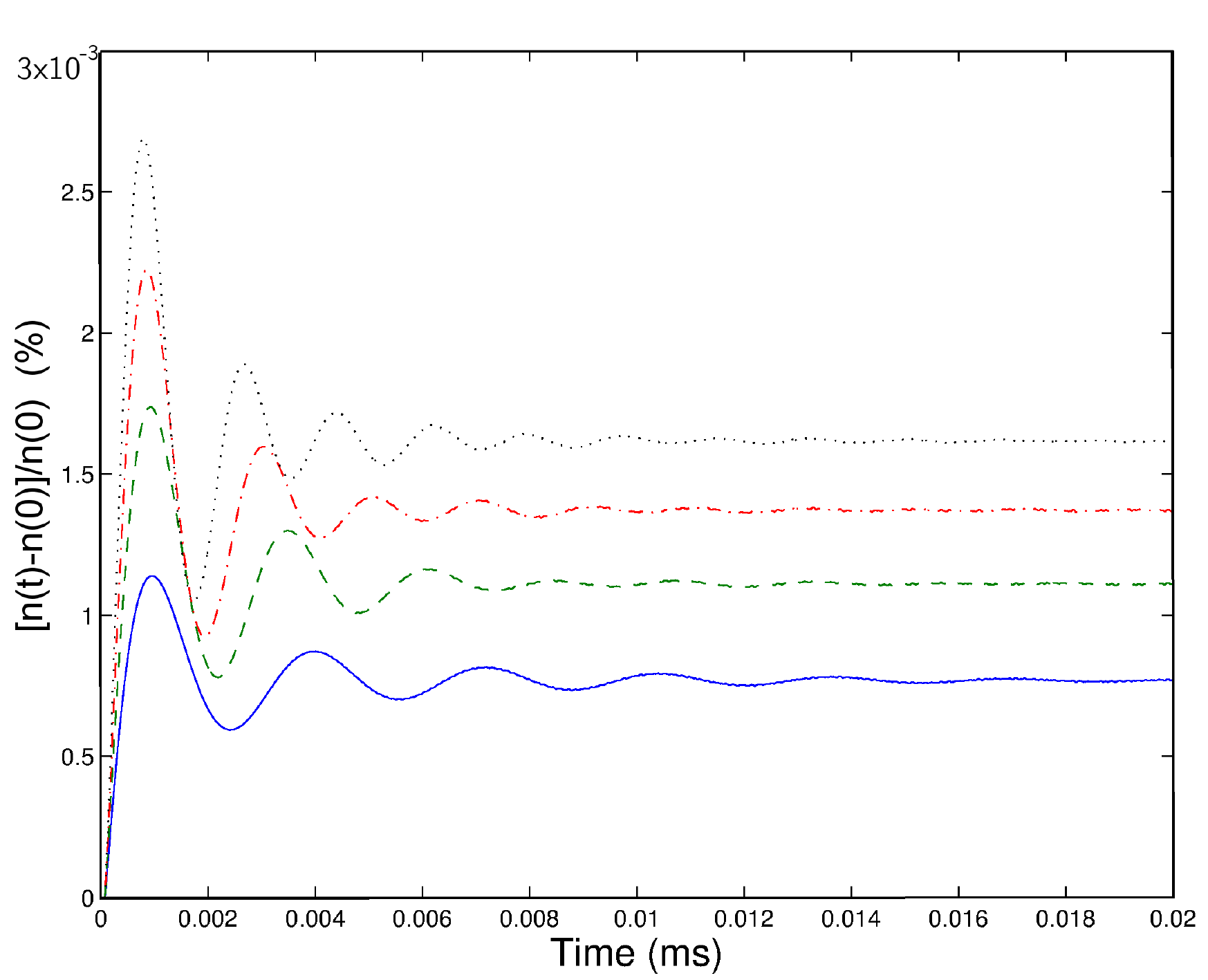}
\caption{(Color online) Evolution of the molecule production efficiency after an infinitely 
fast sweep of the magnetic field across the 198.85~G resonance in
$^{40}\mbox{K}$.
The final field is held constant at 197.5~G. The different lines represent 
different initial magnetic fields of 199~G (solid, blue line), 199.5~G (dashed,
green line), 200~G (dot-dashed, red line) and 200.5~G (dotted, black line). 
Other symbols are as defined in Fig.~\ref{dens_osc_varybf}}.
\label{dens_osc_init_var}
\end{figure}

In Fig.~\ref{dens_osc_init_var} the initial magnetic field is varied and the
final magnetic field held constant. Again the oscillations for all detunings
are on the order of $10^{-3} \%$. Both the frequency and amplitude of the 
oscillations increase as the initial field moves further from the resonance, but
not significantly. It should be noted that this appears to be in contrast to
the $s$-wave where the amplitude increases as the initial field moves towards
the resonance~\cite{Marzena_prl_2005}. However, in the $s$-wave studies
the final field was generally chosen to be much further from the resonance position,
and so corresponds to a BCS like state projected quickly onto a deeply bound BEC
state. The near orthogonality between the initial and final states in such a
case means that oscillations are likely to be small in the mean-field limit. In
our results with the final field closer to the resonance position we see merely
the change in the pairing between initial and final state reflected.
In both cases the
amplitude of the oscillations is very small (the results compared to in the 
$s$-wave correspond to a density of $1.5 \times 10^{13}$~cm$^{-3}$).


\begin{figure}[th]
\centering
 \includegraphics[width=\columnwidth,clip]{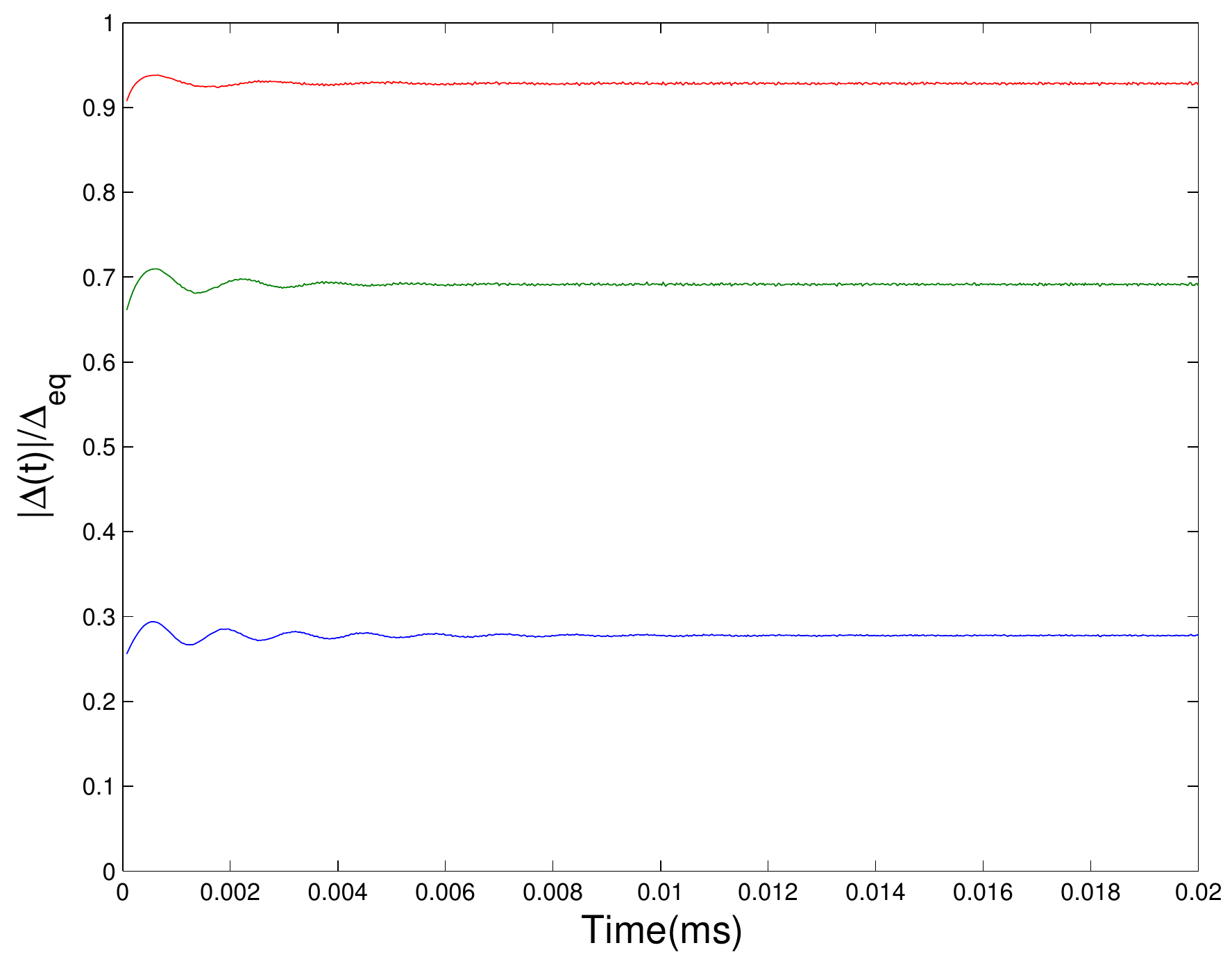}
 \includegraphics[width=\columnwidth,clip]{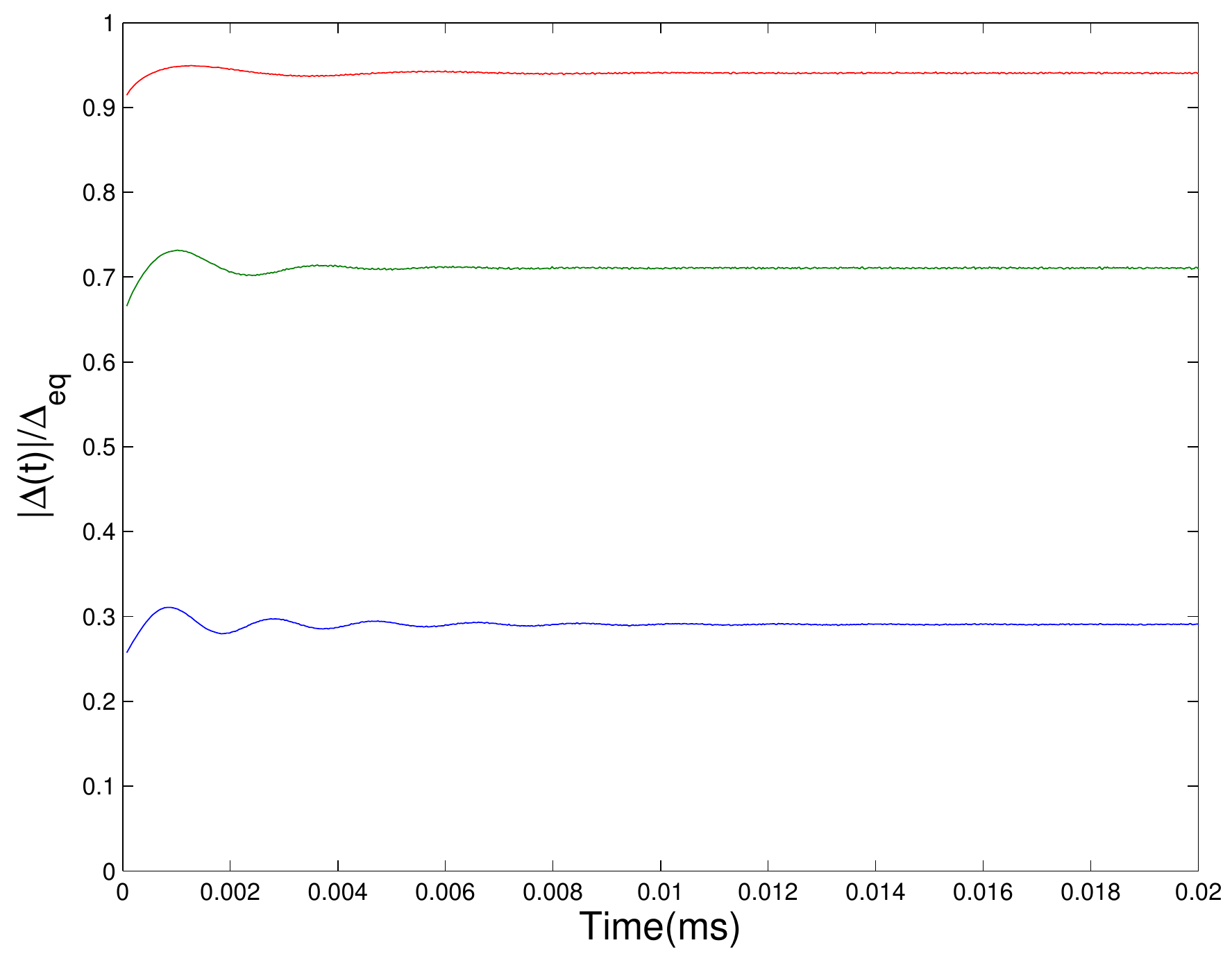}
\caption{(Color online) Variation of the quantity $|\Delta(t)|/\Delta_{eq}$ with time for final
magnetic fields of 197~G (top panel) and 198~G (bottom panel). The different
lines correspond to different initial magnetic field positions of 198.2~G (top,
red), 200.2~G (middle, green) and 201.2~G (bottom, blue).}
\label{gap_osc_init_var}
\end{figure}
We have also studied how the order parameter varies after such a magnetic field
variation. In this case the gap parameter is a function of time defined
by
\begin{equation}
\Delta (t) = \xi \int d^{3} q \langle \chi | \mathbf{q} \rangle \langle \mathbf{q} | \Phi (t) \rangle,
\label{gap_par_td}
\end{equation}
where we have used the separable potential to divide out a form factor from
each side of the equation. We note that the value of the binding energy does
not enter this equation directly. We compare this value against the value
of the gap parameter when the system is in equilibrium at the
final magnetic field position. We note that, in general, the quantity in 
Eq.~(\ref{gap_par_td}) is complex. As for the case of the density variation,
we have studied the effect of varying both the initial and final magnetic fields.

We plot the time evolution of the gap parameter in Fig.~\ref{gap_osc_init_var}. The
top and bottom panels refer to final magnetic fields of $197\mbox{G}$ and
$198\mbox{G}$, respectively. In each panel the different lines correspond to
different initial magnetic field positions of $198.2\mbox{G}$ (top, red),
$200.2\mbox{G}$ (middle, green) and $201.2\mbox{G}$ (bottom, blue). It can be seen
that the closer the initial and final fields are to each other the closer the
value of the gap parameter is to the stationary state value at the final magnetic
field position, denoted here by $\Delta_{eq}$. In all cases, the oscillations have a
small amplitude and quickly decay. The real and imaginary parts of the gap
parameter oscillate at a frequency that corresponds to the energy of the bound 
state at the final magnetic field position. This is expected and serves as a test
on the numerics. The value of oscillation frequency of the absolute value of the
gap parameter increases as the initial magnetic field moves away from the
resonance position and approximately corresponds to the sum of the final bound
state energy and twice the initial chemical potential energy, a measure of the
change in energy of the paired state during the sweep. 

We conclude that it would not be
feasible to observe atom-molecule oscillations in this $p$-wave resonance due
to the small, rapidly decaying amplitude of the density oscillations. 
This is essentially the same conclusion reached in Szyma\ifmmode \acute{n}\else
\'{n}\fi{}ska et al.~\cite{Marzena_prl_2005} but extends this result to the
closed-channel dominated $p$-wave resonance in $^{40}\mbox{K}$. This also suggests
that the method of fast sweeps to probe a fermionic $p$-wave paired
condensate would be a suitable method to probe the condensate were such
conditions favourable.

\section{Discussion}

We have studied the dynamics of $p$-wave Feshbach molecule formation by linear
sweeps of the magnetic field within a mean-field approach.
It turns out that the molecule production efficiency is highly dependent on the
initial state of the gas. 
As the density increases it is increasingly possible to explore a range of
initial magnetic field values, $B_{i}$. As the value of $B_{i}$ moves away
from the resonance and deeper into the BCS region it becomes easier
to produce molecules through a sweep of the magnetic field if the density is
high enough. Moreover, for a given magnetic field value there exists a density
at which molecule production from a linear sweep is optimal. This is due to
the fact that as the density increases so does the value of the molecule 
production from an instantaneous projection of the initial state onto the final
molecular wave function. We have treated the $m_{1}=0$ and $|m_{1}|=1$ 
molecules separately since even at relatively high density the thermodynamic 
states are well described separately. In turn this allows the dynamics to be
treated as if the resonances were uncoupled. The initial assumption that the
partial wave series for the dynamical functions will converge seems to be
justified since adding higher partial waves has little effect on
the production efficiency. We have also shown that, similarly to the
case of an $s$-wave paired system~\cite{Marzena_prl_2005},
it would 
be very difficult to observe atom-molecule coherence as the result of
a fast sweep experiment.

We 
note that our production efficiencies are generally reasonably
high, and in our treatment a sizeable fraction of the atoms can be associated
into molecules using experimentally realistic magnetic ramp speeds. However, we
have not included relaxation processes into our treatment as such terms are
beyond the mean-field theory we have used. Experiments to date on $p$-wave
molecules have found them to be short lived, with lifetimes of between $2$ and
$20~\mbox{ms}$~\cite{zhang,gaebler,fuchs,Inada_2008}, due to dipolar relaxation
processes~\cite{gaebler} and three-body
losses~\cite{levinson_2007,JonaLasinio,Levinson2008}. Our results indicate that
small experimental molecular efficiencies cannot be explained by mean-field
theory.

For single-component Fermi systems with $p$-wave interactions, several
theoretical approaches predict the occurrence of quantum phase
transitions (QPTs) in 2D~\cite{gurarie_2005,Cheng_2005,Botelho2005} and 3D~\cite{Iskin2006} between states with different molecular angular momenta, when the strength the interaction changes. 
The existence of such QPTs might be observable in cold Fermi gases by tuning their interaction through a Feshbach resonance, 
showing particular signatures in the momentum distribution~\cite{Botelho2005,Rombouts2010} or the size of the pairs~\cite{Iskin2006,Lerma2011}; molecule formation might also be affected by the presence of the QPT.
In the case of the resonance studied in this paper, however, the splitting between the
$m=0$ and $m=\pm 1$ resonances is large and appears to indicate that no
QPT will occur under current experimental conditions~\cite{gurarie_radzihovsky_review}, and furthermore that the resonances are approximately independent.
To study molecule formation for other systems in which a QPT may occur, our technique used in this paper might need to be extended to include beyond mean-field terms due to the difficulty in simulating a dynamical
crossing of a QPT. Such an extension could be a generalisation of the approach used by Altman and Vishwanath~\cite{altman_2005} in the $s$-wave case, for example.

The authors acknowledge a significant contribution from T. K\"{o}hler to this work and are grateful to M. Szyma\'{n}ska for useful 
discussions. This work is funded by EPSRC grant EP/E025935/2, Spanish MICINN Project FIS2009-10061, CAM research consortium QUITEMAD, COST Action IOTA (MP1001), and a FP7 Marie Curie fellowship (J.M.-P.).

\appendix
\section{Modelling of the two-body interaction}
\label{app_1}
Previous work has successfully used separable potentials with form factors of a Gaussian form to model 
the $s$-wave two-body interaction in the context of Feshbach resonances
~\cite{Marzena_PRA_2005,Marzena_prl_2005}. This is convenient both 
analytically and numerically because Gaussian integrals can often be
performed analytically.
We wish to describe the system close to threshold and therefore 
choose the form factor to reflect the physics in this region. In the $p$-wave 
we choose the form factor to be
\begin{equation}
\langle \mathbf{p} | \chi_{m_{1}} \rangle = \frac{p \sigma_{m_{1}}}{\pi \hbar^{5/2}}
e^{\frac{-p^{2} \sigma_{m_{1}}^{2}}{2 \hbar^{2}}} Y_{1,m_{1}}(\hat{\mathbf{p}})
\label{form_factor}
\end{equation}
which ensures the correct threshold behaviour of the bound and scattering 
states~\cite{love}. Here $\sigma_{m}$ is a length scale associated with the 
range of the potential and $Y_{1,m_{1}}(\hat{\mathbf{p}})$ is the $\ell=1$ spherical 
harmonic with a projection $m_{1}$ onto the chosen $z$-axis, in this case the 
magnetic field axis.

Using the form factor above and the low energy limit of the scattering 
amplitude it is possible to relate the scattering volume to the parameters
of the separable potential amplitude, $\xi_{m_{1}}$ defined in 
Eq.~(\ref{full_sep_pot_form}), and the form factor, Eq.~(\ref{form_factor}),
\begin{equation}
 a_{1 m_{1}} = 2 \sigma_{m_{1}}^{3} \frac{x}{1 + x/\sqrt{\pi}},
\end{equation}
where $x = m \xi_{m_{1}}/(4 \pi \hbar^2 \sigma_{m_{1}})$ is dimensionless, with
$m$ being the mass of the atom. 
Equation~(\ref{gen_boun_st}) for the bound state energy is then given by
\begin{equation}
 1+\frac{x}{\sqrt{\pi}}\left( 1- 2 y^{2} \left[1-\sqrt{\pi} y e^{y^{2}} 
 \mathrm{erfc}(y)\right]
 \right)= 0,
 \label{psepen}
\end{equation}
where $y = \sigma_{m_{1}} \sqrt{-m E_{-1}}/\hbar$, $E_{-1}~(<0)$ is the bound state 
energy, and $\mathrm{erfc}(y)=\frac{2}{\sqrt{\pi}}\int_y^\infty\exp(-u^2)\,du$ 
is the complementary error function. 

All of our calculations have been performed for the fermionic isotope $^{40}$K.
We have fixed the parameters of our separable potential using the experimental
results of 
Refs.~\cite{ticknor,gaebler}.
To do this, we take the
low energy expansion of the binding energy as given by Eq.~(\ref{penapp}) and 
parameterise the scattering volume as in Eq.~(\ref{resonform}). Close to the
resonance we can expand the resonance formula as a series in powers of 
$(B-B_{\ell,m_{\ell}}^{0})$, the terms of which can be directly read off
from the expression given in Ref.~\cite{ticknor} fixing the resonance 
parameters in Eq.~(\ref{resonform}). The values obtained are given in 
Table~\ref{table_param}.
\begin{table}[hbt]
\begin{center}
\begin{tabular}{c c c c}
 \hline \hline
 $\left| m_{\ell} \right|$ & $B_{\ell,m_{\ell}}^{0}(\mbox{G})$ &
 $\Delta B(\mbox{G})$ & $a^\mathrm{bg}_{\ell m_{\ell}}(a_{\mathrm{B}}^{3})$ \\ \hline
 0 & 198.85 & -21.9482 & -1049850\\
 1 & 198.373 & -24.9922 & -905505\\ 
\hline
\end{tabular}
\end{center}
\caption{Parameters used for modelling the two-body interaction ($a_{\mathrm{B}}$ 
indicates the Bohr radius.)}
\label{table_param}
\end{table}

Equation~(\ref{penapp}) can be differentiated with respect to the magnetic 
field to give the magnetic moment of the molecules relative to the free atoms.
This can be directly equated to the experimental values of the magnetic moment 
allowing the parameter $\sigma_{m_{1}}$ to be calculated. Ref.~\cite{gaebler} 
gives the magnetic moments of the molecules as
\begin{equation}
\left.\frac{\partial E}{\partial B}\right|_{m_{1}=0} = 188 \pm \mbox{2 kHz/G} ,
\end{equation}
and
\begin{equation}
\left.\frac{\partial E}{\partial B}\right|_{|m_{1}|=1} = 193 \pm 2 \mbox{kHz/G} .
\end{equation}
The corresponding range parameter is given by
\begin{equation}
\sigma_{m_\ell} = \frac{m \Delta B_{m_\ell}}{\sqrt{\pi} \hbar^{2}} a_{m_\ell}^{\mathrm{bg}}\left.\frac{\partial
E}{\partial B}\right|_{m_\ell}.
\end{equation}

\bibliographystyle{apsrev}
\bibliography{bib_tex_for_paper}
\end{document}